# Nonlinear dielectric spectroscopy in a fragile plastic crystal

M. Michl, Th. Bauer, P. Lunkenheimer,[a)] and A. Loidl

*Experimental Physics V, Center for Electronic Correlations and Magnetism, University of Augsburg, 86135 Augsburg, Germany*

In this work we provide a thorough examination of the nonlinear dielectric properties of a succinonitrile-glutaronitrile mixture, representing one of the rare example of a plastic crystal with fragile glassy dynamics. The detected alteration of the complex dielectric permittivity under high fields can be explained considering the heterogeneous nature of glassy dynamics and a field-induced variation of entropy. While the first mechanism was also found in structural glass formers, the latter effect seems to be typical for plastic crystals. Morover, the third harmonic component of the dielectric susceptibility is reported, revealing a spectral shape as predicted for cooperative molecular dynamics. In accord with the fragile nature of the glass transition in this plastic crystal, we deduce a relatively strong temperature dependence of the number of correlated molecules, in contrast to the much weaker temperature dependence in plastic-crystalline cyclo-octanol, whose glass transition is of strong nature.

## I. INTRODUCTION

In recent years nonlinear dielectric spectroscopy has proven a useful tool for the investigation of the molecular dynamics in glass forming materials (see, e.g., Refs. 1,2,3,4,5,6,7,8). It was found that the response of glassy matter to high electrical fields goes far beyond the rather trivial polarization saturation effects investigated in some pioneering works on nonlinear dielectric properties (e.g., Refs. 9,10). For example, nonlinear dielectric measurements have provided valuable information on the heterogeneous nature of glassy dynamics[1,2,5,11] and on the role of cooperativity for the slowing down of molecular dynamics at the glass transition.[3,6,12] Both effects are hallmark features of glassy matter, whose thorough exploration is prerequisite for achieving a better understanding of the glass transition and the glassy state of matter in general.[13,14]

Very recently, nonlinear dielectric spectroscopy was also applied to the so-called plastic crystals.[8,15,16] While the molecules in these materials have long-range translational order, their reorientational degrees of freedom exhibit considerable disorder and glassy freezing at low temperatures. Pronounced nontrivial nonlinear behavior was also found in this class of glasslike materials. Interestingly, while some of these effects resemble those found in structural glass formers (i.e., supercooled liquids), in several respects the nonlinear dielectric response of plastic crystals seems to show peculiar differences, despite their linear response behaves very similar to that of conventional glass formers.[17] An important example is the variation of the main relaxational response (the $\alpha$ relaxation) as obtained for high ac fields: In two independent studies,[8,15] the typical relaxation peaks revealed in the frequency dependence of the dielectric loss, $\varepsilon''(\nu)$, were found to exhibit a broadening on *both* flanks when measured at high fields of several 100 kV/cm. In contrast, in supercooled liquids such field-induced broadening was only found at the high-frequency flank of the loss peaks,[1,5] which could be well explained assuming a distribution of relaxation times caused by dynamical heterogeneities.[1,18] Based on measurements of the third-order harmonic component of the susceptibility $\chi_3$, in Ref. 8 indications for molecular cooperativity governing the glassy freezing of plastic-crystalline cyclo-octanol were found. Moreover, in agreement with the behavior of various supercooled liquids,[6] the cooperativity was demonstrated to scale with the apparent temperature-dependent energy barrier (note that according to a recent work[19] anharmonicity may play an addition role). However, in this plastic crystal the detected cooperativity seems to lead to much weaker slowing down of molecular dynamics than in the canonical glass formers.[8] Unfortunately, until now cyclo-octanol seems to be the only plastic crystal were higher harmonics of the susceptibility were reported and further measurements are needed to check for the possible universality of these findings.

In the present work we report a thorough investigation of the nonlinear dielectric behavior of the plastic-crystalline mixture of 60% succinonitrile and 40% glutaronitrile (60SN-40GN). Both, the modification of the dielectric permittivity under high fields and the higher-order susceptibility $\chi_3$ of 60SN-40GN have been investigated. In contrast to pure succinonitrile, this mixture remains in the orientationally disordered state down to low temperatures and crosses over to a glassy-crystal state at an orientational glass temperature of $T_g = 144$ K.[20,21] Most plastic crystals (including cyclo-octanol[22]) can be characterized as rather "strong" glass formers[17,23] within the strong/fragile classification scheme used to account for the different degrees of deviations from Arrhenius behavior of the $\alpha$-relaxation time.[24] Thus 60SN-40GN is an especially

---

[a)]Electronic mail: peter.lunkenheimer@physik.uni-augsburg.de

interesting system as it represents one of the rare examples[25] of a relatively fragile plastic crystal.[20,23,26] Thus it is well suited to check if the relation between apparent energy barrier and $N_{corr}$, suggested for structural glass formers,[6] holds in plastic crystals, too: For a fragile material this relation should lead to a significantly stronger variation of $N_{corr}$ than the relatively weak temperature dependence observed in cyclo-octanol.[8] In addition, we study the field-induced peak broadening, which is also found in 60SN-40GN, however with some deviations from the behavior reported in the other plastic crystals.[8,15] Finally, information on the nonlinear behavior of the two secondary processes in this system is provided.

## II. EXPERIMENTAL DETAILS

Succinonitrile and glutaronitrile with stated purities of ≥ 99% were purchased from Acros Organics and measured without further purification. The mixtures were prepared by putting liquid glutaronitrile into succinonitrile, melted in a water bath, under heavy stirring. The concentration is specified in mol%.

The dielectric experiments were performed using a frequency-response analyzer and the high-voltage booster "HVB 300" from Novocontrol Technologies. The sample material, which is liquid at room temperature, was mixed with 0.1% silica microspheres (2.87 µm average diameter) and put between two highly polished stainless steel plates. The microspheres act as spacing material, ensuring a small plate distance that enables the application of high fields of up to 357 kV/cm.[27] To exclude field-induced heating effects, successive high- and low-field measurements were performed, as described in detail in Ref. 27. For cooling, a closed-cycle refrigerator was used.

## III. RESULTS AND DISCUSSION

### A. Field-induced modification of dielectric permittivity

Figure 1 shows spectra of the dielectric constant $\varepsilon'$ and loss $\varepsilon''$ of plastic-crystalline 60SN-40GN measured at various temperatures. The open symbols represent the results obtained with a low electrical field of $E_l = 13$ kV/cm. They reasonably agree with the previously reported data obtained for 0.1 kV/cm.[20] Obviously, for 13 kV/cm the system is still in the linear regime. The steps in $\varepsilon'(\nu)$ and the main peaks in $\varepsilon''(\nu)$ signify the $\alpha$ relaxation, their continuous shift with temperature reflecting the glassy freezing of molecular reorientation. From the temperature-dependence of the $\alpha$-relaxation time $\tau_\alpha$, related to the peak frequency $\nu_\alpha$ via $\tau_\alpha \approx 1/(2\pi\nu_\alpha)$, an orientational-glass temperature of $T_g = 144$ K was determined.[20] At temperatures below $T_g$, a second peak shifts into the investigated frequency window [Fig. 1(b)] evidencing a secondary relaxation process, denoted as $\gamma$ relaxation.[20] In addition, faint indications for a further relaxation process having characteristic times between those of the $\alpha$ and the $\gamma$ relaxation were obtained in Ref. 20, deduced from a detailed analysis of the spectra.

Based on a criterion by Ngai and Paluch,[28] it most likely represents a Johari-Goldstein $\beta$ relaxation, an intermolecular process thought to be inherent to glassy matter.[29] In Fig. 1(b) the corresponding loss peaks can be assumed to be partly submerged under the $\alpha$-relaxation peaks and to lead to the shallow power law observed, e.g., between about 0.3 and 100 Hz for 141 K.

The closed symbols in Fig. 1 show the results at a high field, $E_h = 357$ kV/cm. A comparison with the low-field spectra (open symbols) reveals clear field-induced variations: While the frequencies of the loss peaks [Fig. 1(b)] seem to be unaffected by the application of a higher field, $\varepsilon''$ rises at both the low and high-frequency flanks of the peaks and the peak widths increase. This finding is in good qualitative agreement with the results for other plastic-crystalline systems reported in Refs. 8 and 15. At high frequencies, when the regime of the mentioned secondary processes is approached, the field-induced increase of $\varepsilon''$ diminishes, again in agreement with earlier measurements.[8,15] The main nonlinear effects for the real part are revealed in the region of the relaxation step where $\varepsilon'$ becomes significantly larger for high fields [Fig. 1(a)]. The latter is also the case for low frequencies, in contrast to cyclo-octanol,[8] where high fields induce a reduction of $\varepsilon'$ at $\nu < \nu_\alpha$ while $\varepsilon'$ increases at $\nu > \nu_\alpha$.

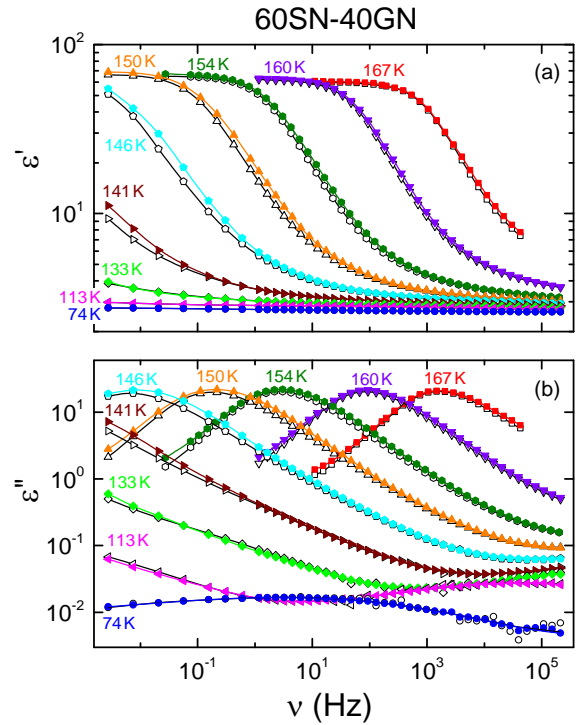

FIG. 1. Dielectric constant (a) and loss spectra (b) of 60SN-40GN at various temperatures. The open and closed symbols show data measured at low (13 kV/cm) and high fields (357 kV/cm), respectively. The lines are guides to the eyes.

To obtain more detailed information on the nonlinear behavior, in Fig. 2 the difference of the high- and low-field spectra is plotted. Following earlier work,[1,5] we show the quantity $\Delta \ln \varepsilon'' = \ln \varepsilon''(E_h) - \ln \varepsilon''(E_l)$. To avoid overcrowding of the figure, in Fig. 2(a) the results at $T \geq 141$ K are shown while those at $T \leq 141$ K are plotted in Fig. 2(b). In Fig. 2(a), the mentioned peak broadening at both flanks of the $\alpha$ peaks corresponds to a V-shaped behavior of $\Delta \ln \varepsilon''$. The minimum is located close to the $\alpha$-peak frequency, which for each temperature is indicated by the arrows in Fig. 2. The behavior revealed by Fig. 2(a) qualitatively resembles that reported in other plastic crystals.[8,15] The increase at $\nu > \nu_\alpha$ can be assumed to have similar origin as in structural glass formers: Within the framework of the so-called box model,[11,30] assuming a distribution of relaxation times caused by dynamical heterogeneities, the field-induced increase of $\varepsilon''$ at $\nu > \nu_\alpha$ arises from a selective transfer of field energy into the heterogeneous regions, accelerating their dynamics.[1,18] In Ref. 15, an explanation for the field-induced increase of $\varepsilon''$ at the low-frequency flank of the $\alpha$ peak of plastic crystals was provided, too: Following theoretical considerations by Johari,[31] based on the Adam-Gibbs theory,[32] it may arise from the reduction of configurational entropy induced by the external field, leading to an increase of the relaxation time. This, however, only affects the data at $\nu < \nu_\alpha$ because the field-induced variation of the entropy is too slow to lead to any significant effect at higher frequencies.[15]

Taking together the results from Refs. 8, 15, and the present work, this entropy-driven nonlinearity seems to be a rather universal property of plastic crystals, while it does not play a significant role in structural glass formers.[1,5,18,27,33] To theoretically deduce this entropy effect, the influence of a high electrical field on the reorientational degrees of freedom of the molecules was considered.[31] It seems reasonable that high fields may diminish the reorientational disorder of dipolar molecules but they should have no effect on the translational disorder. In plastic crystals, reorientational disorder of the molecules provides the main source of entropy. However, in structural glass formers in addition translational degrees of freedom exist and the overall influence of high fields on entropy may be smaller. One may speculate that this is the reason for the found different low-frequency nonlinear behavior of structural glass fomers and plastic crystals.

While the overall nonlinear behavior of 60SN-40GN in the $\alpha$-peak region, discussed in the preceding paragraphs, resembles that in other plastic crystals,[8,15] a closer look at the V-shaped behavior in Fig. 2(a) reveals one characteristic difference: In contrast to the results in Refs. 8 and 15, at the minimum (close to $\nu_\alpha$) $\Delta \ln \varepsilon''$ does not become zero. Moreover, $\Delta \ln \varepsilon'$, shown in Fig. 3(a), also reveals deviations from the same quantity shown for cyclo-octanol in Ref. 8: While a peak occurs at $\nu > \nu_\alpha$ just as in cyclo-octanol, at lower frequencies $\Delta \ln \varepsilon'$ does not become negative, which would be expected for a mere broadening of the relaxation step in $\varepsilon'(\nu)$. Instead $\Delta \ln \varepsilon'$ seems to approach a positive plateau at low frequencies, corresponding to an increase of the static dielectric constant $\varepsilon_s$, which is also revealed by a closer look at Fig. 1(a). Correspondingly the amplitudes of the loss-peaks in Fig. 1(b) also slightly increase for high fields, thus explaining the non-zero value of $\Delta \ln \varepsilon''$ at $\nu_\alpha$. A variation of $\varepsilon_s$ is a known consequence of the trivial saturation effect mentioned in Section I, but in this case the field should induce a reduction instead of the observed increase of $\varepsilon_s$. A possible reason for the observed effect is found when considering that the succinonitrile and glutaronitrile molecules both exist in different conformations, which have different dipolar moments[34,35] For the plastic phase of pure succinonitrile, neutron-scattering experiments explicitly suggest a trans-gauche transition.[34] Thus one may speculate that high electrical fields could induce transitions into conformations with higher dipolar moment. This implies an increase of the average dipolar moment of the sample material and thus explains the observed increase of $\varepsilon_s$. In some respect this scenario resembles the nonlinear effects observed for the so-called Debye process of some monohydroxy alcohols: There stretched conformations of the hydrogen-bonded molecule clusters causing this process[36,37,38] were assumed to be preferred over ring-like structures when a high field is applied.[38,39]

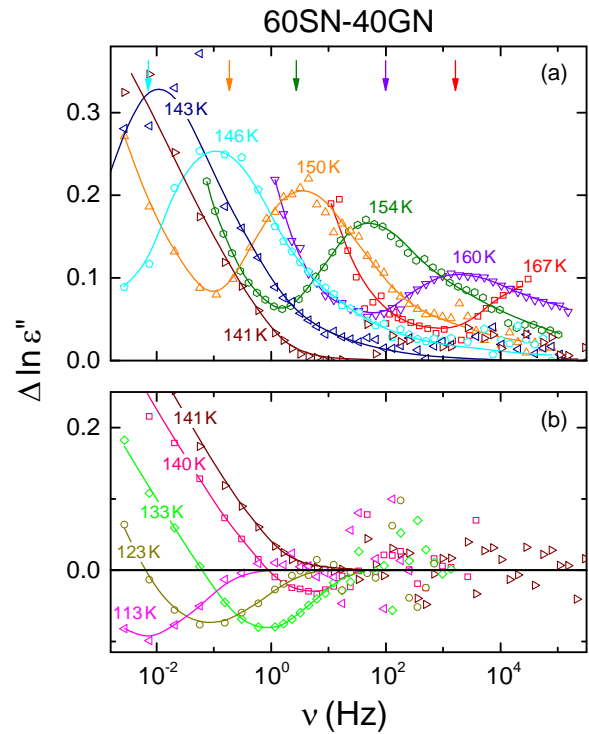

FIG. 2. Difference of the logarithms of the high- and low-field dielectric-loss spectra of plastic-crystalline 60SN-40GN, plotted for various temperatures. The arrows indicate the $\alpha$-peak positions [Fig. 1(b)]. In (a) the data for $T \geq 141$ K and in (b) those for $T \leq 141$ K are shown.

As revealed by Fig. 2(a), about 1.5 frequency decades above the minimum, $\Delta \ln \varepsilon''$ goes through a maximum and starts to decrease again, finally reaching values below the resolution limit of the device (reflected by the strong data scatter). Obviously, at high frequencies, where the response is dominated by the mentioned secondary relaxations, no nonlinear effects are detectable for the temperatures plotted in Fig. 2(a). This reminds of the absence (within experimental resolution) of nonlinear effects in the frequency region of the excess wing or secondary relaxations, previously reported for the structural glass formers glycerol,[5] propylene carbonate,[5] and 1-propanol[27] and for the plastic crystal cyclo-octanol[8] (see Ref. 40 for an explanation of this behavior within the coupling model). $\Delta \ln \varepsilon'$ also approaches zero for high frequencies [Fig. 3(a)]. Interestingly, for the plastic crystals investigated in Ref. 15, a peak in $\Delta \ln \varepsilon''$ followed by a decrease was also observed. However, for the highest frequencies the data reported in Ref. 15 do not approach zero but a plateau is reached. Currently, we have no explanation for this different behavior. In Ref. 41, similar differences found for the excess wing in supercooled liquids[5,41] were proposed to arise from the different numbers of high-field cycles applied to the samples in the different experiments. To check for this effect, additional measurements with a higher number of cycles were performed. For example, at 146 K and frequencies of 1000 Hz and higher, the field was applied for about 5 s, in contrast to less than 1 s for the measurements shown in Fig. 2. Nevertheless, the results were identical for both experiments.

peak, this reduction of $\varepsilon''$ and $\varepsilon'$ reflects the behavior at $\nu > \nu_\beta$. It either corresponds to a reduction of relaxation strength, which could be caused by polarization saturation, or could reflect an increase of the $\beta$-relaxation time under high field. In both cases the behavior does not agree with that reported for the $\beta$ relaxation of the structural glass former sorbitol, where the amplitude of the secondary mode was found to increase for high fields.[42] In the region of the $\gamma$ relaxation, no significant field-induced variation is revealed by Figs. 2(b) and 3(b) but one should be aware that, at least for the loss, the uncertainty of the data is rather high for high frequencies. Overall, the nonlinearity in the secondary-relaxation regime of 60SN-40GN is clearly weaker than in the $\alpha$ regime, but the details of the field-induced variation, especially of the Johari-Goldstein type $\beta$ relaxation which is strongly superimposed by the $\alpha$ relaxation, are difficult to resolve.

## B. Third-order nonlinear dielectric susceptibility

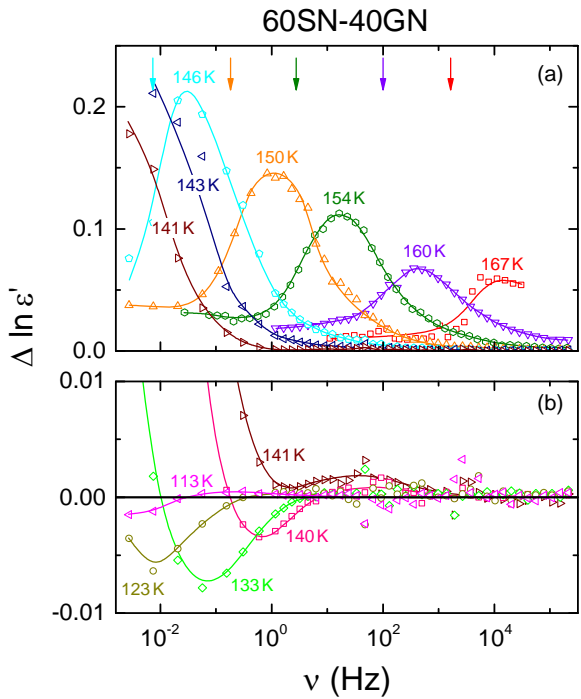

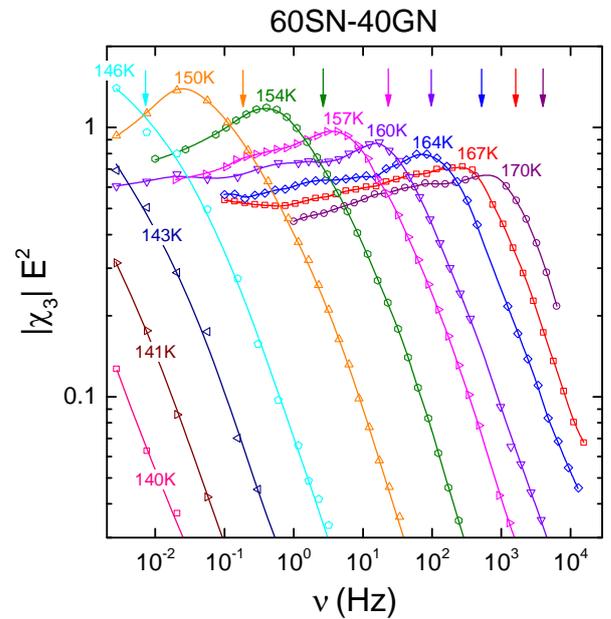

FIG. 3. Difference of the logarithms of the high- and low-field dielectric-constant spectra of plastic-crystalline 60SN-40GN, plotted for various temperatures. The arrows indicate the $\alpha$-peak positions [Fig. 1(b)]. In (a) the data for $T \geq 141$ K and in (b) those for $T \leq 141$ K are shown.

FIG. 4. Third-order harmonic component of the dielectric susceptibility of plastic-crystalline 60SN-40GN. Spectra of $|\chi_3|E^2$ are shown for various temperatures, measured at a field of 357 kV/cm. The arrows indicate the $\alpha$-peak positions. The lines are guides to the eyes.

While Figs. 2(a) and 3(a) do not reveal any nonlinear behavior in the regime of the secondary relaxations, for the lowest investigated temperatures, $T < 141$ K, nevertheless some nonlinearity is found [Figs. 2(b) and 3(b)]. Here negative values are detected and only after going through a minimum, $\Delta \ln \varepsilon'$ and $\Delta \ln \varepsilon''$ finally approach zero. This minimum seems to occur in the region of the $\beta$ relaxation, which leads to the shallow power law showing up, e.g., in the region around 1 Hz for the 133 K curve in Fig. 1(b). At low temperatures, the three relaxations of 60SN-40GN discussed above become more separated[20] and, obviously, only then this nonlinearity of the $\beta$ relaxation is revealed. As the left flank of the $\beta$ relaxation peak still is submerged under the $\alpha$

Figure 4 shows spectra of the dimensionless quantity $|\chi_3|E^2$ for 60SN-40GN at various temperatures. The third-order harmonic component of the susceptibility $\chi_3$, accessible by nonlinear dielectric measurements, recently has proven an important quantity to learn more about the cooperative nature of glassy dynamics. Within the model by Biroli, Bouchaud, and coworkers,[12,43] cooperativity was predicted to lead to a pronounced hump in $|\chi_3|(\nu)$, which indeed was found in several canonical glass formers.[3,6] (However, it should be noted that a hump in $|\chi_3|(\nu)$ may also arise within the framework of other models.[44,45,46,47])

Moreover, as mentioned in Section I, in plastic-crystalline cyclo-octanol such behavior was observed, too. As revealed by Fig. 4, 60SN-40GN also shows a hump in $|\chi_3|(\nu)$, located nearly one decade below the $\alpha$-peak frequency, which is indicated by the arrows in Fig. 4 for each temperature. Within the theoretical framework of Refs. 12 and 43, this indicates that cooperativity also plays a role for the glassy dynamics in this plastic crystal.

A closer inspection of Fig. 4 reveals weak shoulders in the $|\chi_3|$ spectra, about one decade below the hump frequencies. Currently, it is not clear what causes this spectral feature. Interestingly, in Ref. 20, based on excess intensity in the loss detected below the $\alpha$ peak frequency, indications for an additional process, slower than the $\alpha$ relaxation were found. Thus one may speculate that the observed shoulders in Fig. 4 are associated with this process, whose microscopic origin is unknown until now.

From $|\chi_3|$, the quantity $X = |\chi_3| k_B T / [(\Delta\varepsilon)^2 V \varepsilon_0]$ can be calculated, also showing a hump[3,12,48] ($\Delta\varepsilon$ is the relaxation strength, $V$ the volume taken up by a single molecule, and $\varepsilon_0$ the permittivity of free space). Its amplitude $X_{max}$ should be approximately proportional to the number of correlated molecules $N_{corr}$.[3,12,48] In Fig. 5(b), $N_{corr}(T)$ of 60SN-40GN, estimated in this way, is shown (pentagons). It reveals a significant increase with decreasing temperature. Its temperature variation is clearly stronger than for the only other plastic crystal investigated in this way, cyclo-octanol,[8] whose $N_{corr}(T)$ is shown by the inverted triangles in Fig. 4. As mentioned in Section I, 60SN-40GN represents a rare example of a plastic crystal with a relatively fragile temperature characteristics of its glassy freezing, while cyclo-octanol exhibits rather strong behavior. This becomes obvious by a comparison of their temperature-dependent $\alpha$-relaxation times in the Angell plot[49] ($\tau_\alpha$ vs. $T_g/T$) shown in Fig. 5(a). While $\tau_\alpha(T)$ of cyclo-octanol nearly follows Arrhenius behavior (nearly linear behavior in Fig. 5),[22] SNGN shows clear deviations.[20] Correspondingly, the fragility index[50] of 60SN-40GN ($m = 62$)[20] is significantly larger than for cyclo-octanol ($m = 33$)[22]. The temperature ranges where $N_{corr}$ of these two plastic crystals was investigated in the present work and in Ref. 8 is indicated by the horizontal bars in Fig. 5(a). Obviously, for the more fragile 60SN-40GN the temperature dependence of $\tau_\alpha$ in this range is significantly stronger than for cyclo-octanol. When assuming that cooperativity causes the non-Arrhenius behavior of glassy matter, the stronger temperature dependence of $N_{corr}$ as revealed in Fig. 5(b) is thus fully consistent with the higher fragility of 60SN-40GN.

Moreover, for several canonical glass formers[6] [glycerol, propylene carbonate (PCA), 3-fluoroaniline (FAN), and 2-ethyl-1-hexanol (2E1H)] and for plastic-crystalline cyclo-octanol,[8] a direct proportionality of $N_{corr}$ and the effective energy barrier of the $\alpha$ relaxation were found. The latter can be estimated from the derivative in an Arrhenius plot of the $\alpha$-relaxation time, $H = d(\ln\tau)/d(1/T)$.[6,51] This proportionality is demonstrated in Fig. 5 by the reasonable agreement of the $H(T)$ and scaled $N_{corr}(T)$ curves (lines and symbols, respectively) shown for the structural glass formers investigated in Ref. 6 and for the plastic crystal cyclo-octanol.[8] In the present work, we find the proportionality $H \propto N_{corr}$ to be rather well fulfilled for 60SN-40GN, too [cf. the pentagons and the corresponding line in Fig. 5(b)].

For the canonical glass formers glycerol, PCA, and FAN, the scaling factor $a$ in Fig. 5 (used to make the $N_{corr}(T)$ and $H(T)$ curves match) was found to be of the order of one, varying between 0.72 and 1.3.[6] (2E1H has $a = 0.39$ but it represents a special case where the observed relaxation is due to the formation of molecular clusters.[6,36]) In contrast, as pointed out in Ref. 8, for cyclo-octanol a significantly smaller value of $a = 0.19$ was found, which implies that in this plastic crystal the same $N_{corr}$ leads to a lesser impediment of the molecular motions than in supercooled liquids. This was attributed to different intermolecular coupling mechanisms in plastic crystals compared to supercooled liquids. However, in the plastic crystal 60SN-40GN investigated in the present work, we find a scaling factor close to unity ($a = 1.05$), just as for the supercooled liquids. It seems likely, that this different behavior compared to cyclo-octanol is related to the fact that this mixed system has strong substitutional disorder. Moreover, the strongly different conformations of the two molecule species forming this system may also play a role. Clearly there are more degrees of freedom in 60SN-40GN than in the pure plastic crystal cyclo-octanol, which can also be said for the supercooled liquids and, thus, may explain their similar behavior. However, the detailed microscopic mechanism behind this finding is unclear at present and more experiments on plastic crystals are necessary to clarify this issue.

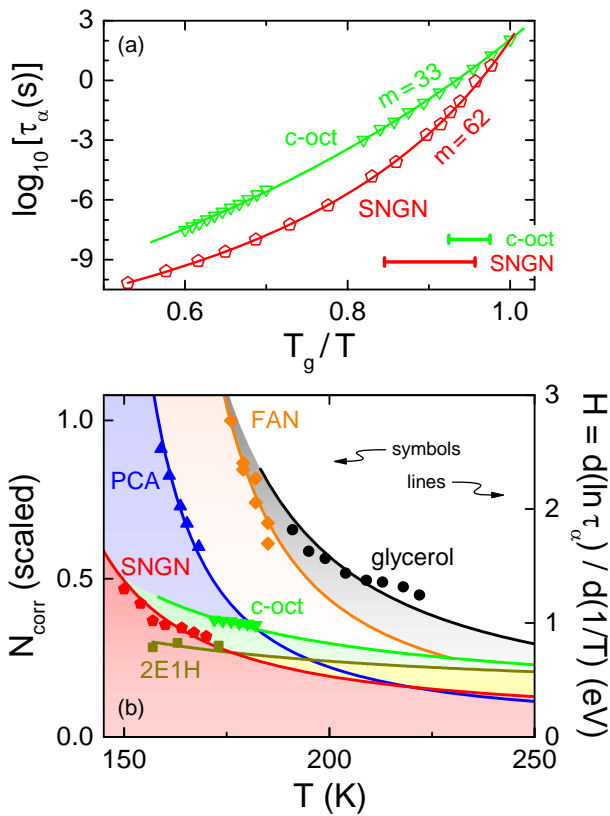

FIG. 5. (a) Angell plot of the $\alpha$-relaxation time of cyclo-octanol[22] and 60SN-40GN.[20] The temperature ranges (converted into $T_g/T$) where $N_{corr}$ of these two substances was investigated are indicated by the horizontal bars. (b) Comparison of activation energies and $N_{corr}$. The lines show the effective activation energies $H$ of plastic-crystalline 60SN-40GN (present work, pentagons, abbreviated as SNGN in the figure) and cyclo-octanol[8] (c-oct) and of various structural glass formers[6] determined from the derivatives of their temperature-dependent relaxation-times (right scale). For the same materials, the symbols show the number of correlated molecules $N_{corr}$ (left scale). $N_{corr}$ was determined from $\chi_3$ and is shown in arbitrary units. The $N_{corr}$ data were multiplied by separate factors $a$ for each material [glycerol: 1.15, PCA: 0.72, FAN: 1.30, 2E1H: 0.39, cyclo-octanol: 0.19, 60SN-40GN: 1.05] leading to a good match with the derivative curves. Note that both ordinates in Fig. 4(b) start from zero, implying direct proportionality of both quantities.

## V. SUMMARY AND CONCLUSIONS

In summary, in this work we have provided a detailed investigation of the nonlinear dielectric properties of the mixed plastic-crystalline system 60SN-40GN by measuring both the field-induced modification of the complex permittivity and the third-order harmonic component of the susceptibility. The first type of nonlinear measurement reveals a broadening of the loss peaks, associated with the $\alpha$-relaxation, at high fields. This broadening occurs both at low and high frequencies, in marked contrast to canonical glass formers, where only the high-frequency flank of the loss peaks is affected by the high field. This effect is in good agreement with the findings in other plastic crystals reported by our group[8] and by Richert and coworkers[15] and thus seems to be a universal property of this class of glassy matter. The high-frequency effect can be ascribed to the heterogeneous nature of glassy dynamics,[1,18] which obviously is present in plastic crystals just as in supercooled liquids. The low-frequency effect most likely arises from a field-induced reduction of configurational entropy as proposed in Ref. 15.

Moreover, the high-field permittivity of 60SN-40GN exhibits some peculiar features, not found in the other plastic crystals: High applied fields lead to a significant increase of the static dielectric constant, which most likely can be ascribed to a field-induced rise in molecular conformations with higher dipolar moments. In addition, at low temperatures the loss caused by the $\beta$ relaxation becomes reduced by high fields, most likely indicating an increase of the $\beta$-relaxation time. In contrast, for the observed $\gamma$ relaxation no nonlinear effect was deteced, similar to our findings for the secondary relaxations in various structural glass formers[5,27] and in plastic-crystalline cyclo-octanol.[8]

In addition, the third-harmonic susceptibility of 60SN-40GN was measured, to our knowledge representing only the second case where this quantity was investigated for a plastic crystal. Just as for plastic-crystalline cyclo-octanol[8], a clearly humped shape of the spectra of $|\chi_3|$ is found as theoretically predicted for cooperative glassy dynamics by Biroli and coworkers.[12,43] Determining the number of correlated molecules $N_{corr}$ from its amplitude reveals a significant increase with decreasing temperature and a significantly stronger temperature variation than in cyclo-octanol. When considering the higher fragility of 60SN-40GN compared to cyclo-octanol, this finding is consistent with a cooperativity-induced origin of the non-Arrhenius behavior of glassy dynamics. Moreover, just as in various canonical glass formers and in cyclo-octanol, a direct proportionality of the effective energy barrier and $N_{corr}$ was found, similar to the assumptions made within the Adam-Gibbs theory of the glass transition.[32] Indeed this seems to be a rather universal property of glassy matter, now also affirmed in a strong (see Ref. 8) and a fragile plastic crystal.

## ACKNOWLEDGMENTS

This work was partly supported by the Deutsche Forschungsgemeinschaft via Research Unit FOR 1394.